\documentclass[preprint,12pt,authoryear]{elsarticle}
\usepackage[a4paper, top=2.5cm, bottom=2.5cm, left=2cm, right=2cm]{geometry}
\usepackage{amssymb}
\usepackage{amsmath}
\usepackage{multirow}%
\usepackage{amsmath,amssymb,amsfonts}%
\usepackage{amsthm}%
\usepackage{mathrsfs}%
\usepackage[title]{appendix}%
\usepackage{xcolor}%
\usepackage{textcomp}%
\usepackage{manyfoot}%
\usepackage{booktabs}%
\usepackage{algorithm}%
\usepackage{algorithmicx}%
\usepackage{algpseudocode}%
\usepackage{listings}%
\usepackage{natbib}
\usepackage{booktabs}
\usepackage{bm}
\usepackage{hyperref}
\usepackage[T1]{fontenc}    
\usepackage[utf8]{inputenc} 
\usepackage{subcaption}
\usepackage{booktabs} 
\usepackage{tabularx} 
\usepackage{threeparttable}
\usepackage{caption}  

\def\be{\begin{equation}}
	\def\ee{\end{equation}}
\def\bea{\begin{eqnarray}}
	\def\eea{\end{eqnarray}}
\journal{High Energy Astrophysics}

\graphicspath{{Graphics/}}

\def\be{\begin{equation}}
	\def\ee{\end{equation}}
\def\bea{\begin{eqnarray}}
	\def\eea{\end{eqnarray}}

\begin{document}
	
	\begin{frontmatter}

		\title
		{Cosmological Constraints on the Phenomenological
			Interacting Dark Energy Model with \emph{Fermi} Gamma-Ray Bursts and DESI DR2}
		
		\author[label1]{Ziyan Zhu\fnref{email1}}
		\author[label2]{Qingquan Jiang\fnref{email2}}
		\author[label3]{Yu Liu\fnref{email3}}
		\author[label4]{Puxun Wu\fnref{cor4}}
		\author[label1]{Nan Liang\corref{cor1}}
		
		\affiliation[label1]{organization={Guizhou Key Laboratory of Advanced Computing},
			addressline={Guizhou Normal University},
			city={Guiyang},
			postcode={550025},
			state={Guizhou},
			country={China}}
		\affiliation[label2]{organization={School of Physics and Astronomy},
			addressline={China West Normal University},
			city={Nanchong},
			postcode={637009},
			country={China}}
		\affiliation[label3]{organization={School of Physical Science and Technology},
			addressline={Southwest Jiaotong University},
			city={Chengdu},
			postcode={611756},
			state={Sichuan},
			country={China}}
		
		\affiliation[label4]{organization={Department of Physics and Synergistic Innovation Center for Quantum Effects and Applications},
			addressline={Hunan Normal University},
			city={Changsha},
			postcode={410081},
			country={China}}
		\cortext[cor1]{Corresponding author: liangn@bnu.edu.cn}
		\cortext[cor2]{Corresponding author: qqjiangphys@yeah.net}
		\cortext[cor3]{Corresponding author: lyu@swjtu.edu.cn}
		\cortext[cor4]{Corresponding author: pxwu@hunnu.edu.cn}
		\fntext[email1]{zzy@gznu.edu.cn}
		
		
		
		%
		
\begin{abstract}
In this work, we constrain the phenomenological interacting dark energy (IDE) model using \emph{Fermi} gamma-ray burst (GRB) dataset and the latest baryon acoustic oscillation (BAO) data from the Dark Energy Spectroscopic Instrument (DESI)  Data Release 2 (DR2). Through a joint Bayesian analysis, we perform a cosmological comparative assessment of the $\Lambda$CDM, $w$CDM, and CPL models with the phenomenological IDE model. For the phenomenological IDE model in a flat universe with \emph{Fermi} samples and DESI DR2, we obtain: $\xi=2.63^{+0.63}_{-0.52}$,  $\xi + 3w_X = -0.98^{+1.90}_{-2.07}$ with the GOLD sample ($1.4\le z \le5.6$)  and $\xi=2.83^{+0.63}_{-0.58}$,   $\xi + 3w_X = 0.03^{+1.35}_{-1.33}$ with the FULL sample ($1.4\le z \le8.2$), respectively. Our analysis shows that the $\Lambda$CDM model without interaction ($\xi=3$,  $\xi + 3w_X = 0$) is consistent with the latest \emph{Fermi} sample and DESI DR2 at $1\sigma$ confidence level. We find no significant deviations from the standard model using AIC and BIC criterias.
\end{abstract}
		
		
	\end{frontmatter}
	
	\section{INTRODUCTION}
	
	The discovery of the universe's accelerated expansion via Type Ia supernovae \citep{Riess1998, Perlmutter1999} cemented the $\Lambda$CDM model as the standard cosmological paradigm, which is plagued by fundamental theoretical puzzles, including the fine-tuning and coincidence problems \citep{Weinberg1989, Carroll2001}.
Moreover, increasingly precise observations have revealed 
the Hubble tension \citep{Di Valentino2025}: a discrepancy of $H_0$  at $\gtrsim 5\sigma$
between the local $H_0$ value \citep{Riess2022} and that inferred from Cosmic Microwave Background (CMB) 
\citep{Planck2020}. 
The observational data in the redshift region between the local distance ladder calibrated by Cepheids ($z<0.01$) and CMB ($z\sim1000$) might offer important insights into the origins of the $H_0$ tension.
Recently, the Dark Energy Spectroscopic Instrument (DESI) 
release the baryon acoustic oscillation (BAO) \citep{DESI2024}, which suggest 
that dark energy (DE) may not be a cosmological constant \citep{Colgain2026}.

These challenges motivate the exploration of dynamical DE models and extensions to the standard model 
\citep{Feng2005, Copeland2006, Liang2009, Liang2011, WangLiang2010, Li2013, Li2019, Li2020, Liu2022a, WLL2024, Liu2025a, Liu2025b, Gupta2023, Lei2025, Das2024, Fazzari2025}.
The Chevallier-Polarski-Linder (CPL) parameterizations \citep{Chevallier2001, Linder2003} provides a wide framework for testing time-varying the equation of state (EoS): $w(z) = w_0 + w_a z/(1+z)$, where $w_0$ represents the present value and $w_a$ characterizes its time evolution.
For new physics beyond the standard model, the interacting dark energy (IDE) models 
	have garnered significant attention by introducing a non-gravitational interaction between (dark) matter and DE, 
in which the energy flow transfer from DE to matter can alleviate the coincidence problem \citep{Amendola2000, Zimdahl2001, Wang2019, Costa2014, Yang2021}. 
In particular, 
the interaction between  DE and (dark) matter
can effectively raise the theoretically predicted value of $H_0$, bringing it closer to local observational results. 
Therefore, the IDE models can potentially reconcile observational discrepancies to alleviate the Hubble tension \citep{Pan2019, Pan2020, Kumar2019, DiValentino2021, Abdalla2022}.
Moreover, it is interesting to find that the energy flow transfer between DE and DM with sign shifting interaction functions \citep{Halder2024, Silva2025, Li2026}.

The phenomenological scenario can be explored the IDE model by considering the ratio of the energy densities of dark energy to matter 
as \citep{Dalal2001} $r \equiv \rho_{X} \propto \rho_{m}a^{\xi}$, where $\rho_{\rm X}$ and $\rho_{\rm m}$ are the energy densities of dark energy and matter, respectively. 
The cases where $\xi = 3$ and $\xi = 0$ correspond to the $\Lambda$CDM model and a self-similar solution without the coincidence problem, respectively; values of $0 \leq \xi \leq 3$ indicate a less severe coincidence problem. 
In a flat FRW (Friedmann-Robertson-Walker) universe, the interaction term can be obtained by $Q=-H\rho_m(\xi+3w_X)$, where $w_{\rm X}$ represents the EoS for DE. The condition $\xi+3w_X\neq 0$ denotes the interacting scenario.
Some early studies investigated the phenomenological model with the joint data utilizing SNe Ia, CMB, as well as the BAOs and the observational Hubble data (OHD), which indicated that the coincidence problem was not alleviated \citep{Guo2007,Chen2010,Cao2011}.
However, many works found that the latest data 
suggest energy transfer from DE to DM \citep{Wang2022,Goswami2025}.
Other observational data have also been used to constrain the phenomenological IDE models, e.g.,
strong gravitational lensing (SGL)  data \citep{Cheng2021},
mock data of gravitational wave (GW) \citep{Zheng2022,Hou2023,Li2024a}, as well as the mock and localized Fast radio bursts (FRBs) \citep{Zhao2023,Yan2025}.  

Gamma-ray bursts (GRBs) are the most intense explosions observed to reach a higher maximum observable redshift at $z \sim 10$ \citep{Cucchiara2011}. 
The luminosity 
relations of GRBs  with known redshift between measurable spectroscopic properties 
and their luminosity or energy have been proposed to investigate cosmology \citep{Amati2002, Ghirlanda2004, Yonetoku2004, Dai2004, LZ2005, LZ2006, Schaefer2007, Dainotti2008}.
In order to avoid the circularity problem, \cite{Liang2008} introduced a cosmological model-independent method for calibrating the luminosity relations of 69 GRBs \citep{Schaefer2007} 
from SNe Ia.
On the other hand, \cite{Amati2008} also proposed the simultaneous fitting method which constrain the coefficients of the relationship and the parameters of the cosmological model simultaneously. 
Furthermore, \cite{Amati2019} proposed an alternative approach for calibrating the Amati relation  of  193 GRBs with firmly measured redshift and spectral parameters taken from \cite{Demianski2017} 
using OHD from the cosmic chronometers (CC) method
by simultaneous fitting
with the B\'ezier parametric. 

Therefore, GRBs data can be employed to place constraints on cosmological models without the circularity problem \citep{Kodama2008, Liang2010, Liang2011, Wei2010, Gao2012, Pan2013, Wang2016AA, Wang2019, Montiel2021, Khadka2021, Luongo2021, Liu2022b, Liang2022, Li2023, Luongo2023, WangLiang2024, WLL2024, NongLiang2024, Cao2024, Cao2025, Xie2025a, Zhang2025, Huang2025a, Huang2025b, Bargiacchi2025, Alfano2025, Luongo2025a, Luongo2025b, Luongo2025}.
\cite{Khadka2021} made a compilation 
with the Amati relation \citep{Amati2002} comprising 118 GRBs (A118) with the smallest intrinsic dispersion from a broader set of 220 GRBs (A220).
\cite{Liang2022} calibrated the A219 sample \footnote{Removed GRB051109A, which are counted twice in the A220 sample \citep{Khadka2021}. } from the Pantheon sample \citep{Scolnic2018} 
by using a Gaussian Process to constrain DE models with GRBs at high redshift.

Recently,
\cite{NongLiang2024} tested the phenomenological IDE model with Pantheon+ SNe Ia and A118 GRB dataset.
\cite{WLL2024} investigate the phenomenologically emergent dark energy (PEDE) model \citep{Li2019} and the Generalised Emergent Dark Energy (GEDE) \citep{Li2020} with A118 sample and OHD at intermediate redshift.
\cite{Li2024b,Li2025a} utilize the DESI Data Release 1 (DR1) in combination with SNe Ia from the full five-year observations of the Dark Energy Survey
and  CMB  from the Planck satellite to explore potential interactions between DE and DM.
\cite{Li2026} investigated sign-changeable IDE models using DESI DR1, CMB, and DES SN data to find strong evidence for energy transfer direction reversal during cosmic evolution at the $4.1\sigma$ confidence level for certain interaction forms.

The latest measurements from DESI Data Release 2 (DR2) \citep{DESI2025} provided exquisite distance constraints at redshifts $0.295 < z_{\rm eff} < 2.330$.
\cite{Ling2025} adopted an improved inverse distance ladder approach based on  the latest cosmological data with DESI DR2, cosmic chronometer (CC), and SN Ia from either the DESY5 or Pantheon+ datasets.
Recent analyses using DESI data have provided new constraints on the phenomenological IDE models \citep{Giare2024,Sabogal2024,Silva2025}.
Other alternative DE models beyond $\Lambda$CDM have also been explored with DESI data, e.g., the PEDE model \citep{HA2024} and the GEDE model \citep{Liu2025a}, the torsion cosmology \citep{Liu2025b}, the generalized Padé expansions \citep{Fazzari2025} and dynamical DE parametrization models \citep{Li2025b, Cheng2025}.

More recently,
\cite{WangLiang2024} presented a sample of long GRBs  from 15 years of the \emph{Fermi}-GBM (Gamma-ray Burst Monitor) catalogue with identified redshift, in which the GOLD sample contains 123 long GRBs at $z\le5.6$ and the FULL sample contains 151 long GRBs with redshifts at $z\le8.2$.
\cite{Cao2025} analyzed these \emph{Fermi} GRBs datasets to constrain the Amati relation and cosmological parameters within six 
DE models simultaneously.
\cite{Xie2025b}  tested the Distance Duality Relation with cosmological observations at high redshift combined the \emph{Fermi} sample using artificial neural network (ANN).

In this paper, we investigate the specific phenomenological IDE model 
with the updated sample from the \emph{Fermi} satellite \citep{WangLiang2024} 
and	
the latest BAO measurements from DESI DR2 \citep{DESI2025}. 
This approach allows us to place robust constraints on the interaction strength $\xi$ and the EoS parameter $w_X$, as well as to test the time evolution of dark energy through the CPL parameterization, thereby extending the cosmological probe to higher redshifts and breaking degeneracies among DE models.


\section{PHENOMENOLOGICAL INTERACTING DARK ENERGY MODEL}


The phenomenological IDE model serves as a useful parameterization for several theoretically motivated scenarios, which
introduces a coupling term $Q$ in the continuity equations, ensuring total energy-momentum conservation \citep{Amendola2000}
\begin{align}
\dot{\rho}_m + 3H\rho_m &= Q, \\
\dot{\rho}_X + 3H(1+w_X)\rho_X &= -Q,
\end{align}
where $\rho_m$ and $\rho_X$ are the energy densities of dark matter and dark energy, respectively, $w_X$ is the dark energy equation-of-state parameter, $H \equiv \dot{a}/a$ is the Hubble parameter, and overdots denote derivatives with respect to cosmic time. The sign convention for $Q$ follows: $Q > 0$ implies energy transfer from dark energy to dark matter.

In this work, we use $\xi$IDE to represent the phenomenological interaction model \citep{Dalal2001}, 
with the energy density ratio defined as:
\begin{equation}
\label{eq:density_ratio}
\frac{\rho_X}{\rho_m} \propto a^{\xi},
\end{equation}
where $\xi$ quantifies the interaction strength. This ansatz leads to the interaction function:
\begin{equation}
Q = -H\rho_m(\xi + 3w_X)\Omega_X,
\end{equation}
where $\Omega_X \equiv \rho_X/(\rho_m + \rho_X)$ is the DE density fraction. The sign of the combination $(\xi + 3w_X)$ dictates the direction of energy flow: negative values of $(\xi + 3w_X)$  corresponds to $Q > 0$ with a transfer of energy from DE to DM. 

The background evolution is governed by the Friedmann equation:
\begin{equation}
H^2 = \frac{8\pi G}{3}(\rho_m + \rho_X),
\end{equation}
with the explicit solution for the Hubble parameter in a flat FRW universe:
\begin{equation}
\label{eq:hubble}
\begin{split}
H(z) = H_0 & \left[(1-\Omega_{m0})(1+z)^{3(1+w_X) + \xi} + \Omega_{m0}(1+z)^3\right]^{1/2} \\
& \times \left[\frac{\Omega_{m0} + (1-\Omega_{m0})(1+z)^{-\xi}}{1 + (1-\Omega_{m0})/\Omega_{m0}}\right]^{-3w_X/(2\xi)}.
\end{split}
\end{equation}

\section{DATA AND METHODOLOGY}
In this work, we employ the updated sample of GRBs from the \emph{Fermi} satellite \citep{WangLiang2024}, in which the GOLD sample contains 123 GRBs at $z\le5.6$ and the FULL sample contains 151 GRBs with redshifts at $z\le8.2$.
the latest BAO measurements from the second data release (DR2) of the DESI \citep{DESI2025}, which  provide precise measurements 
across a redshift range of $0.295 < z_{\rm eff} < 2.330$. 

The Amati relation \citep{Amati2002}, which correlates the spectral peak energy $E_{\rm p,\rm i}$ and the isotropic-equivalent radiated energy $E_{\rm iso}$:
\begin{equation}
\log_{10} \left( \frac{E_{\rm iso}}{\text{1 erg}} \right) = a + b \log_{10} \left( \frac{E_{\rm p,\rm i}}{\text{300 keV}} \right).
\end{equation}
The \emph{Fermi}-GBM catalogue at low-redshift can be calibrated from the latest OHD  with the cosmic chronometers method by using a Gaussian Process to obtained GRBs at high-redshift $z\ge1.4$. The corresponding $\chi^2$ function is defined as:
\begin{equation}
\label{eq:chi_grb}
\chi^2_{\mathrm{GRB}} = \sum_{i=1}^{N_{\mathrm{GRB}}}
\frac{[\mu_{\mathrm{obs}}(z_i) - \mu_{\mathrm{th}}(z_i; \mathbf{p}, H_0)]^2}{\sigma_{\mu,i}^2},
\end{equation}
where $\mu_{\mathrm{obs}}$ and $\mu_{\mathrm{th}}$ are the observed and theoretical distance moduli, and $\mathbf{p}$ represents the cosmological parameters, $N_{\mathrm{GRB}}$ represents the Number of \emph{Fermi} GRBs at $z\ge1.4$: for the GOLD sample, $N_{\mathrm{GRB}}=60$; 
and for the FULL sample, $N_{\mathrm{GRB}}=78$, 
respectively.

The BAO observable is expressed as:
\begin{equation}
\label{eq:chi_bao}
\chi^2_{\mathrm{BAO}} = (\mathbf{D}_{\mathrm{obs}} - \mathbf{D}_{\mathrm{th}})^{T} \mathbf{C}_{\mathrm{BAO}}^{-1} (\mathbf{D}_{\mathrm{obs}} - \mathbf{D}_{\mathrm{th}}),
\end{equation}
where $\mathbf{D} = [D_M(z)/r_d, H(z) r_d]$, $D_M(z)$ is the comoving angular diameter distance, $H(z)$ is the Hubble parameter, and $r_d$ is the sound horizon \footnote{The sound horizon $r_d$ is computed self-consistently for each cosmological model by integrating the sound speed in the baryon-photon fluid over the modified expansion history $H(z)$:
$
r_d = \int_{z_d}^\infty \frac{c_s(z)}{H(z)}dz,
$
where $z_d$ is the redshift at the drag epoch and $c_s(z)$ is the sound speed in the baryon-photon fluid.} at the drag epoch.
The sound horizon can be  fixed to fiducial values \citep{Huang2025a} or treated as a free parameter 
\citep{LDX2020}. 
In this work, we adopt $r_{\mathrm{d}}=147.05\pm 0.30\,\ \rm{Mpc}$ \citep{Planck2020} and do not use external $H_0$ priors.\footnote{It should be noted that there is a strong degeneracy between the sound horizon $r_d$ and Hubble constant $H_0$ \citep{Liu2025c}. 
\cite{Liu2024} proposed a model-independent geometric method using low-redshift 
SNe Ia combined with BAO to calibrate the relative standard ruler. 
\cite{Yang2025} determined the Hubble constant and sound horizon from late-time Universe observations by  fully model-independent cosmographic approaches using Taylor series or Padé polynomials.}
We utilize the full covariance matrix from DESI DR2 \footnote{\url{https://github.com/CobayaSampler/bao_data/tree/master/desi_bao_dr2}}, which includes correlations between different redshift bins and between $D_M(z)/r_d$ and $H(z)r_d$ measurements.




The total likelihood combines BAO and GRB datasets:
\begin{equation}
\mathcal{L}_{\text{total}} = \mathcal{L}_{\text{BAO}} \times \mathcal{L}_{\text{GRB}},
\end{equation}
with the corresponding $\chi^2$ function:
\begin{equation}
\label{eq:chi_total}
\chi^2_{\mathrm{total}} = \chi^2_{\mathrm{GRB}} + \chi^2_{\mathrm{BAO}}.
\end{equation}

Our cosmological analysis encompasses four DE models:
\begin{itemize}
\item $\Lambda$CDM: The standard cosmological model with $\Omega_m$ and $H_0$ as free parameters;
\item $w$CDM: Extension with constant dark energy equation of state $w$ as an additional parameter;
\item CPL: Dynamical dark energy with time-varying equation of state $w(z) = w_0 + w_a z/(1+z)$;
\item $\xi$IDE: Interacting dark energy with coupling parameter $\xi$ and equation of state $w_X$.
\end{itemize}

For the Markov Chain Monte Carlo analysis, we employ the \textbf{emcee} Python package \citep{ForemanMackey2013}. Parameter inference and uncertainty quantification are performed using its affine-invariant ensemble sampler.
Posterior distributions obtained from both configurations are in excellent agreement, demonstrating the stability of the sampling and the reliability of our results. Throughout the analysis, all cosmological parameters are assigned physically motivated priors.


\section{RESULTS}
\begin{figure}[H]
\centering
\includegraphics[width=0.35\textwidth]{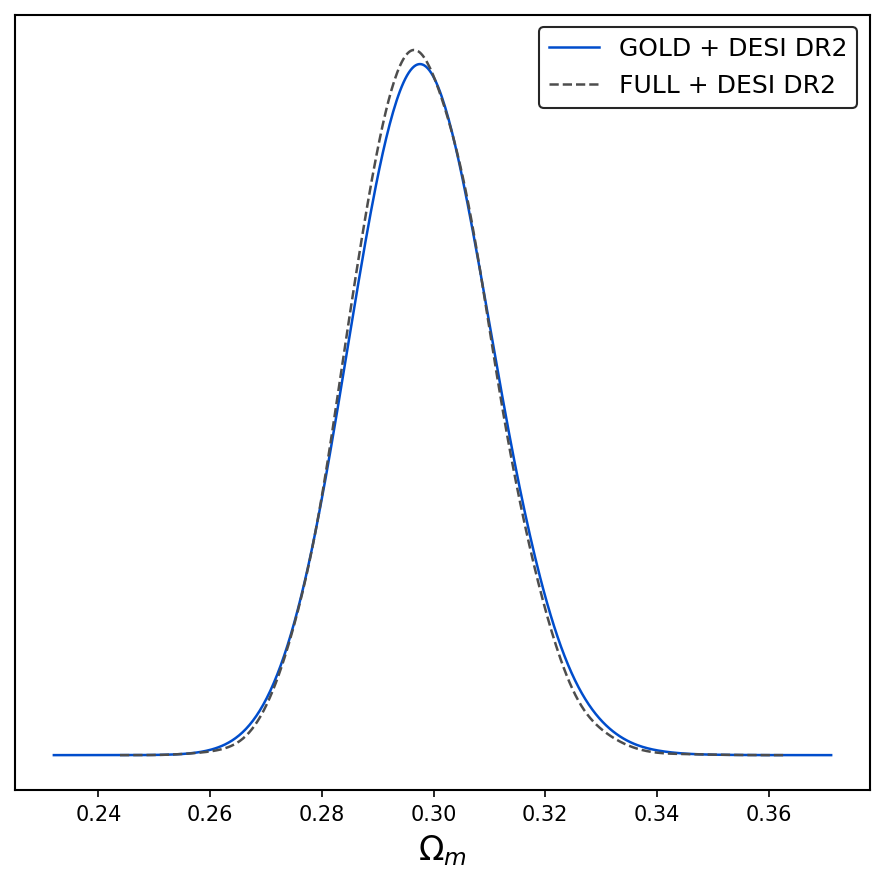}
\hfill
\includegraphics[width=0.45\textwidth]{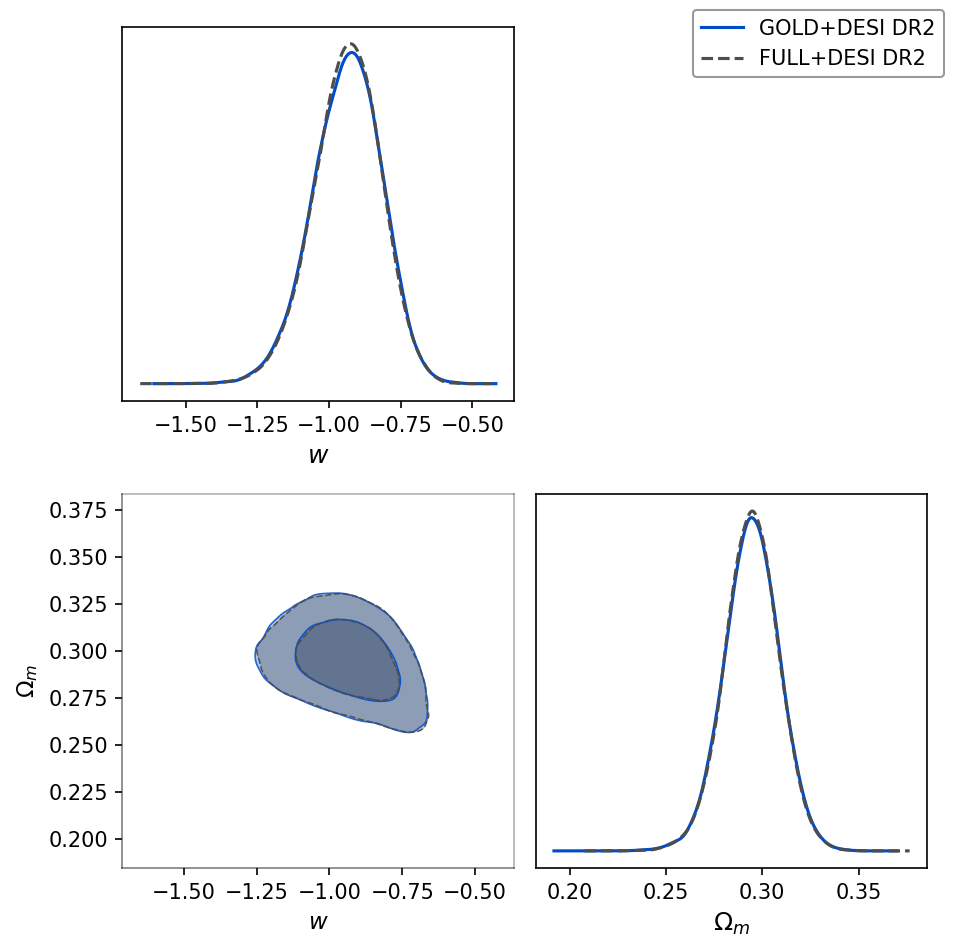}
\caption{Constraints on cosmological parameters (68\% and 95\% confidence levels) from the GOLD GRB dataset (blue) and FULL GRB dataset (gray) for $\Lambda$CDM model (left) and $w$CDM model (right).}
\label{fig:lcdm_wcdm}
\end{figure}

\begin{figure}[H]
\centering
\includegraphics[width=0.61\textwidth]{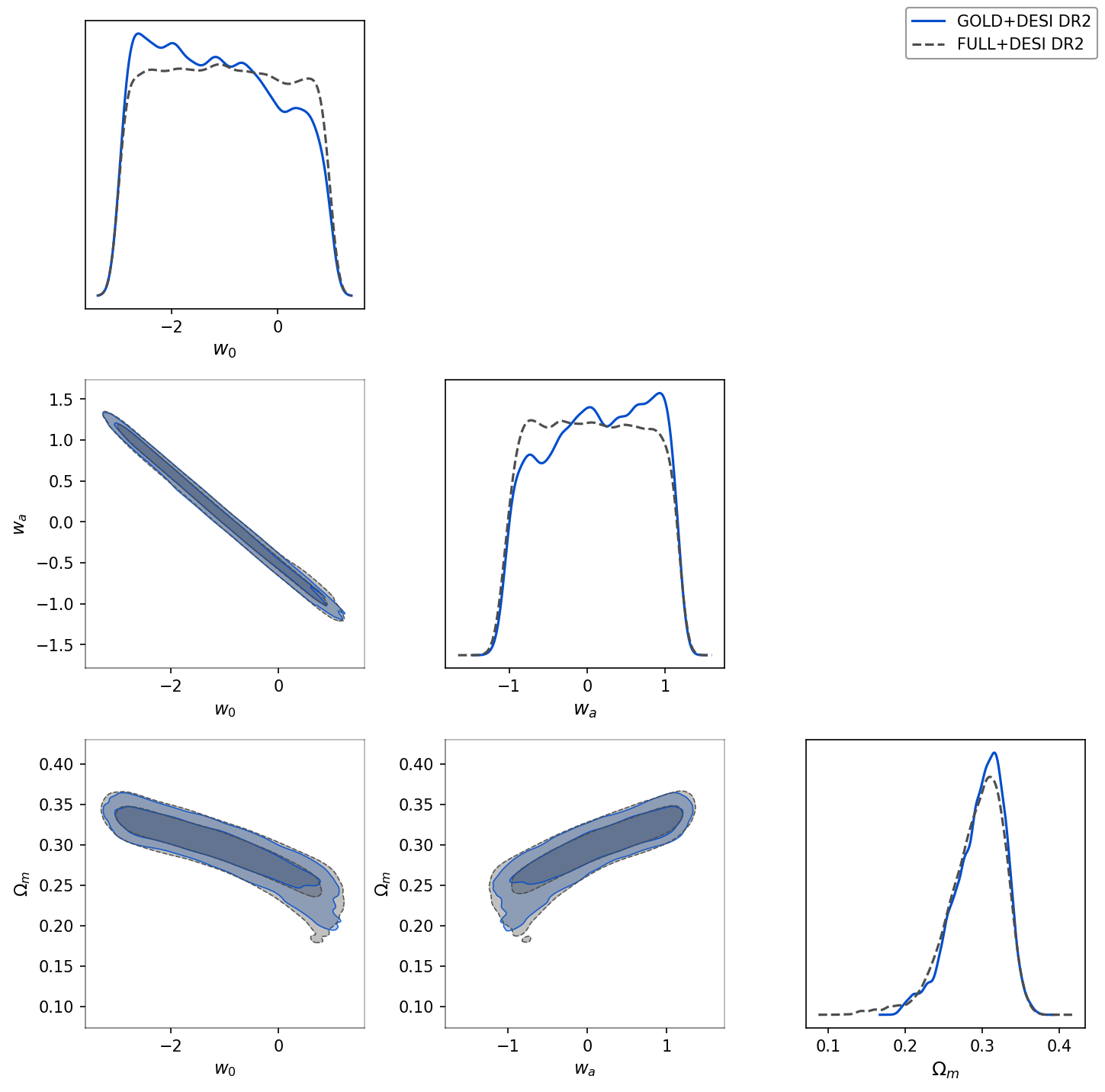}
\caption{Constraints on cosmological parameters (68\% and 95\% confidence levels) from the GOLD GRB dataset (blue) and FULL GRB dataset (gray) for CPL model.}
\label{fig:cpl_single}
\end{figure}

\begin{figure}[H]
\centering
\includegraphics[width=0.61\textwidth]{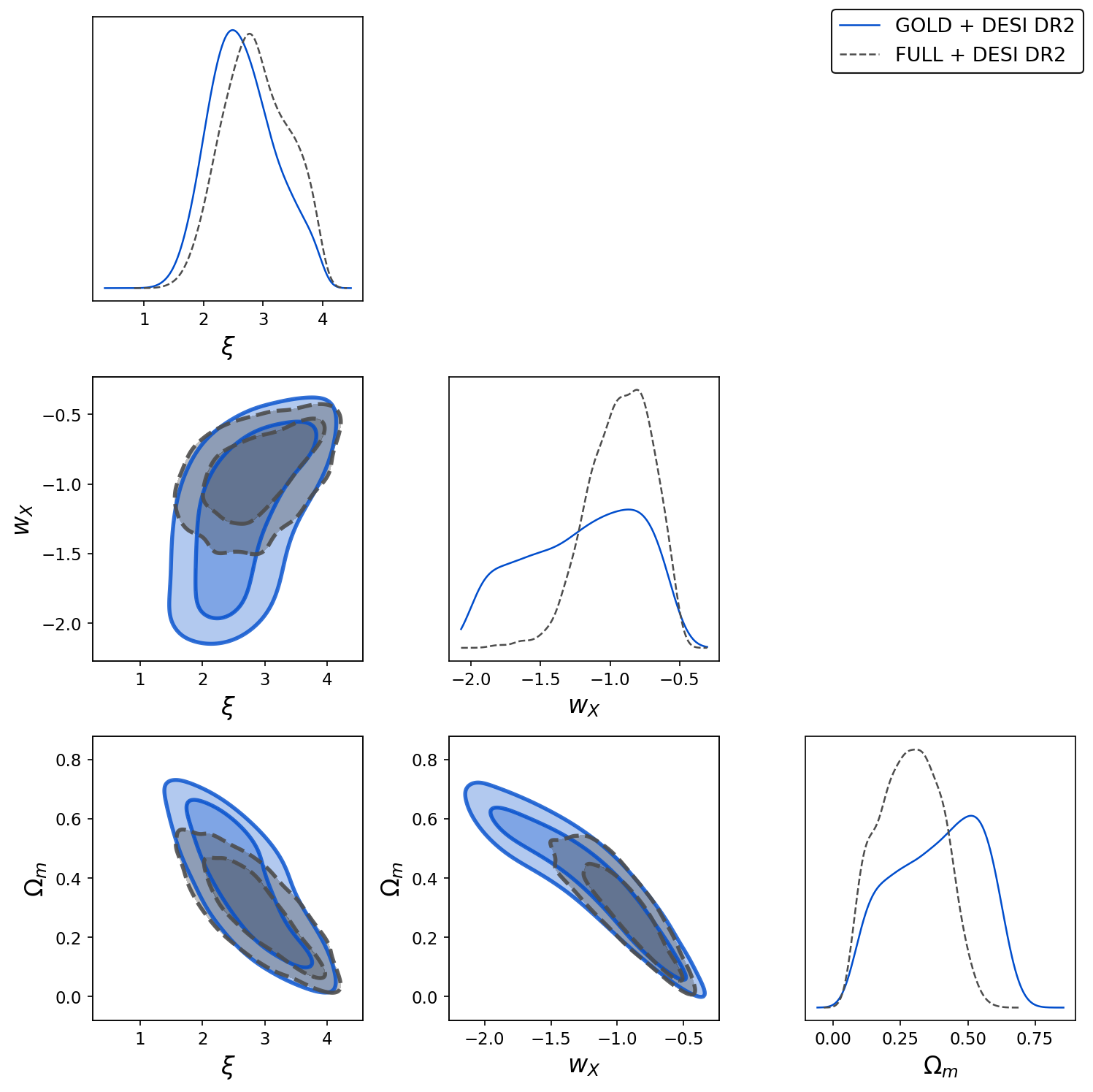}
\caption{Constraints on cosmological parameters (68\% and 95\% confidence levels) from the GOLD GRB dataset (blue) and FULL GRB dataset (gray) for $\xi$IDE model.}
\label{fig:ide_single}
\end{figure}

The joint analysis of DESI DR2 measurements and \emph{Fermi} GRBs provides tight constraints on cosmological parameters while allowing for robust model discrimination. We show constraints from the GOLD and FULL sample  at $z \geq 1.4$ with DESI DR2 BAO measurements for $\Lambda$CDM and $w$CDM models in Figure \ref{fig:lcdm_wcdm}, and  CPL and $\xi$IDE  models in Figure \ref{fig:cpl_single} and \ref{fig:ide_single}; which facilitating direct comparison between the GOLD and FULL samples at high redshifts.
We summarize results of the four DE models ($\Lambda$CDM, $w$CDM,  CPL and $\xi$IDE  models) with parameter constraints and model selection statistics in Table \ref{tab:cosmological_constraints_combined}.
\begin{table*}[h]
\centering
\scriptsize
\renewcommand{\arraystretch}{1.8}   
\setlength{\tabcolsep}{4pt}
\caption{Comparative cosmological constraints at $1\sigma$ confidence level from the GOLD and FULL sample at $z \geq 1.4$  combined with DESI BAO measurements.}
\label{tab:cosmological_constraints_combined}
{
	\scriptsize 
	\setlength{\tabcolsep}{3pt}
	\begin{tabular}{l c c c c c c c c c c}
		\hline\hline
		Dataset & Model & $H_0$ & $\Omega_m$ & $w$/$w_0$ & $w_a$ & $w_X$ & $\xi$ & $-2\ln\mathcal{L}_{\rm max}$ & $\Delta$AIC & $\Delta$BIC \\
		\hline
		\multirow{4}{*}{GOLD+BAO}
		& $\Lambda$CDM & $68.98 \pm 0.54$ & $0.298^{+0.012}_{-0.013}$ & -- & -- & -- & -- & 197.320 & 0.000 & 0.000 \\
		& $w$CDM & $68.09^{+1.60}_{-1.39}$ & $0.292\pm 0.011$ & $-0.928^{+0.110}_{-0.107}$ & -- & -- & -- & 205.322 & 7.212 & 7.161 \\
		& CPL & $67.81^{+2.2}_{-1.9}$ & $0.301^{+0.026}_{-0.038}$ & $-1.134^{+1.400}_{-1.243}$ & $0.134^{+0.720}_{-0.767}$ & -- & -- & 213.452 & 15.123 & 29.112 \\
		& $\xi$IDE & $68.29 \pm 1.7$ & $0.408^{+1.462}_{-0.101}$ & -- & -- & $-1.202^{+0.421}_{-0.518}$ & $2.631^{+0.631}_{-0.516}$ & 206.342 & 11.172 & 31.332 \\
		\hline
		\multirow{4}{*}{FULL+BAO}
		& $\Lambda$CDM & $69.02 \pm 0.53$ & $0.300^{+0.011}_{-0.012}$ & -- & -- & -- & -- & 196.332 & 0.000 & 0.000 \\
		& $w$CDM & $67.96^{+1.69}_{-1.24}$ & $0.297^{+0.014}_{-0.017}$ & $-0.937^{+0.117}_{-0.215}$ & -- & -- & -- & 204.132 & 7.536 & 11.560 \\
		& CPL & $67.88^{+2.23}_{-2.05}$ & $0.294^{+0.032}_{-0.043}$ & $-0.931^{+1.376}_{-1.396}$ & $0.002^{+0.789}_{-0.738}$ & -- & -- & 212.207 & 13.576 & 35.783 \\
		& $\xi$IDE & $67.64 ^{+1.67}_{-1.42}$ & $0.300^{+0.112}_{-0.152}$ & -- & -- & $-0.932^{+0.241}_{-0.250}$ & $2.830^{+0.631}_{-0.582}$ & 209.423 & 13.614 & 28.630 \\
		\hline\hline
	\end{tabular}
}
\begin{tablenotes}
\item \small \textit{Note.} 
$\Delta$AIC and $\Delta$BIC are calculated relative to the $\Lambda$CDM model for each dataset. 
\end{tablenotes}
\end{table*}

From Figure \ref{fig:lcdm_wcdm}, \ref{fig:cpl_single}, \ref{fig:ide_single} and Table \ref{tab:cosmological_constraints_combined}, we find that the results with GOLD and FULL sample are consistent within $1\sigma$ uncertainties. 
For the $\Lambda$CDM model, the results from the two GRB samples are fully consistent and in excellent agreement with Planck CMB measurements \citep{Planck2020}. 
For the $w$CDM model, the constraints obtained from both samples are consistent with the cosmological constant value ($w = -1$) at the $1\sigma$ confidence level. 
For the CPL model, 
the values of $w_a$ are consistent with zero within its $1\sigma$ uncertainty, which suggest that current data cannot support the time evolution of the DE EoS.
For the $\xi$IDE model, 
the energy flow parameter $(\xi + 3w_X)$ with both samples shows sample dependence and remains consistent with no energy exchange within $1\sigma$ uncertainties, indicating no statistically significant preference for any particular transfer direction \citep{Cai2005, Clemson2012, Borges2008}.
The estimated parameters of three extended models 
remain consistent with the expectations of $\Lambda$CDM within $1\sigma$ uncertainties, which show that the $\Lambda$CDM remains a good fit to the combined analysis of high-redshift GRB data and DESI DR2 BAO measurements.

We also compare these DE models with Akaike Information Criterion (AIC), Bayesian Information Criterion (BIC).
The $\Delta\text{AIC}$ and $\Delta\text{BIC}$ values of the four cosmological models are summarized  in Table \ref{tab:cosmological_constraints_combined}
with the $\Lambda$CDM model as the reference, 
which indicate that the 
CPL and $\xi$IDE models have essentially no empirical support compared to the $\Lambda$CDM model.

\begin{table*}[h]
\centering
\scriptsize
\renewcommand{\arraystretch}{1.6}   
\setlength{\tabcolsep}{4pt}
\begin{threeparttable}
\caption{Comparison of constraints for the IDE models at $1\sigma$ confidence level from different studies.}
\label{tab:comparison}
\begin{tabular}{l >{\centering\arraybackslash}p{2.8cm} >{\centering\arraybackslash}p{2.3cm} >{\centering\arraybackslash}p{2.4cm} >{\centering\arraybackslash}p{2.4cm} >{\raggedleft\arraybackslash}p{1.6cm}}
	\hline
	Study & Dataset 
	& $w_X$ & $\xi$ & $\xi + 3w_X$ \\
	\hline
	This work  & GOLD+BAO  
	& $-1.20^{+0.42}_{-0.52}$ & $2.63^{+0.63}_{-0.52}$ & $-0.97^{+1.89}_{-2.07}$ \\
	& FULL+BAO  
	& $-0.93^{+0.24}_{-0.25}$ & $2.83^{+0.63}_{-0.58}$ & $0.03^{+1.35}_{-1.33}$ \\
	\hline
	\citet{NongLiang2024} &  A118+Pantheon+ 
	& $-1.02^{+0.16}_{-0.18}$ & $0.54 \pm 0.36$ & $-2.52 \pm 0.87$ \\
	\hline
	\citet{Zheng2022}
	& GW (Mock) 
	& $-1.24^{+0.38}_{-0.41}$ & $3.05^{+0.77}_{-0.45}$ & $-0.67^{+1.79}_{-1.58}$ \\
	\hline
	\citet{Yang2021} & CMB+BAO+RSD 
	& $-1.02 \pm 0.03$ & $0.08 \pm 0.07$ & $-2.98 \pm 0.08$ \\
	\hline
	\citet{Wang2016} & CMB+BAO+SN 
	& $-1.10 \pm 0.10$ & $0.20 \pm 0.30$ & $-3.10 \pm 0.40$ \\
	\hline
	\citet{Salvatelli2014} & CMB+SN+$H_0$ 
	& $-0.98 \pm 0.02$ & $-0.60^{+0.12}_{-0.13}$ & $-3.54^{+0.12}_{-0.13}$ \\
	\hline
	\citet{Cao2011}
	& $H(z)$+BAO+CMB 
	& $-1.11^{+0.38}_{-0.43}$ & $3.35^{+1.37}_{-1.32}$ & $0.02^{+2.51}_{-2.61}$ \\
	\hline
	\citet{Chen2010}
	& SNe+BAO+CMB 
	& $-0.98 \pm 0.07$ & $3.06 \pm 0.35$ & $0.12 \pm 0.56$ \\
	\hline
	\citet{Guo2007}
	& SNLS+BAO+CMB 
	& $-1.03 ^{+0.12}_{-0.15}$ 
	($2\sigma$) & $3.36  ^{+0.69}_{-0.70}$ 
	($2\sigma$) & $0.27 ^{+1.05}_{-1.15}$ 
	($2\sigma$) \\
	\hline
\end{tabular}
\begin{tablenotes}
\item \small \textit{Note.} 
In this work, our results show two constraints using GOLD (60 GRBs) and FULL (78 GRBs) samples at $z \geq 1.4$ with DESI DR2.
\end{tablenotes}
\end{threeparttable}
\end{table*}

Many previous studies \citep{Guo2007,Chen2010,Cao2011,Salvatelli2014,Wang2016,Yang2021,Zheng2022,NongLiang2024} have tested the phenomenological IDE model with A118 and Pantheon+,  the mock GW data, combinations of SNe (the Supernova Legacy Survey, SNLS), BAO, CMB and Redshift Space Distortion (RSD) data.
Constraints on the parameters of the $\xi$IDE models compared with previous studies are summarised in Table \ref{tab:comparison}.
Compared to the results in the early studies $\xi>3$ \citep{Guo2007,Chen2010,Cao2011,Zheng2022} which indicated that the coincidence problem was not alleviated, and $\xi$ close to 0 \citep{Salvatelli2014,Wang2016,Yang2021} which correspond to a self-similar solution without the coincidence problem, 
our analysis with high-redshift GRBs and DESI DR2 find the values of $\xi<3$, which indicate that the coincidence problem can be alleviated;
the values of $\xi$ close to 3 and $(\xi + 3w_X)$ to 0, which are consistent with the $\Lambda$CDM model without interaction.

\section{CONCLUSION AND DISCUSSIONS}
In this work, we constrain the phenomenological IDE model using \emph{Fermi} GRB dataset  with the Amati relation and the latest BAO data from DESI DR2.
Our comprehensive analysis 
with the $\xi$IDE model shows statistically significant non-zero interaction strength in both samples ( $\xi=2.631^{+0.631}_{-0.516}$,  $\xi + 3w_X = -0.975^{+1.894}_{-2.070}$  for GOLD and $\xi=2.830^{+0.631}_{-0.582}$,   $\xi + 3w_X = 0.034^{+1.354}_{-1.332}$ for FULL), 
which are consistent with the $\Lambda$CDM model without energy exchange ($\xi + 3w_X = 0$) within $1\sigma$ uncertainties.
For the CPL parameterization, which shows large uncertainties in the dark energy evolution parameters, we find $w_a$ consistent with zero at $1\sigma$ confidence level, indicating no evidence for time-varying dark energy equation of state.
Furthermore, none of the considered models significantly alleviates the Hubble tension, with all yielding $H_0$ values consistent with early-universe measurements.
We also find no significant evidence supporting deviations from the standard model by the $\Delta\text{AIC}$ and $\Delta\text{BIC}$ values.

Our analysis considers only one specific phenomenological coupling form and a particular parameterization for dynamical dark energy; other theoretically motivated interaction functions and dark energy parameterizations may yield different conclusions. The cosmological constraints at high-redshift come primarily from GRBs, 
which are sensitive to dark sector interactions with larger systematic uncertainties than low-redshift probes, are not included in our purely geometric analysis.

In conclusion, 
both the IDE scenario and dynamical DE models remain theoretically motivated and observationally testable frameworks that warrant continued investigation with future precision datasets, which 
will dramatically improve cosmological 
constraints by providing 
better-calibrated GRB samples. The French-Chinese satellite spacebased multi-band astronomical variable objects monitor (SVOM) 
will allow for a more stringent test of the phenomenological framework.

\section*{Declaration of competing interest}
The authors declare that they have no known competing financial interests or personal relationships that could have appeared to influence the work reported in this paper.

\section*{ACKNOWLEDGMENTS}
We thank the anonymous referees for their helpful comments and constructive suggestions.
This project was supported by the Guizhou Provincial Science and Technology Foundation: QKHJC-ZK[2021] Key 020  and QKHJC-ZK[2024] general 443.
Q. Jiang was supported by the Sichuan Provincial Key Projects for Science-Education Integration (25LHJJ0097).
Y. Liu was supported by the NSFC under Grant No. 12373063.
P. Wu was supported by the National Natural Science Foundation of China (Grant No. 12275080).

\section*{DATA AVAILABILITY}
Data are available at the following references:
the DESI DR2 BAO data from \cite{DESI2025},
the \emph{Fermi} gamma-ray burst sample from \cite{WangLiang2024}.


\label{lastpage}
\end{document}